\documentclass[journal]{IEEEtranTIE}
\usepackage{graphicx}
\usepackage{cite}
\usepackage{picinpar}
\usepackage{amsmath}
\usepackage{url}
\usepackage{flushend}
\usepackage[latin1]{inputenc}
\usepackage{colortbl}
\usepackage{soul}
\usepackage{multirow}
\usepackage{pifont}
\usepackage{color}
\usepackage{alltt}
\usepackage[hidelinks]{hyperref}
\usepackage{enumerate}
\usepackage{breakurl}
\usepackage{epstopdf}
\usepackage{pbox}
\usepackage{amsfonts}
\usepackage{subcaption}

\newcommand{\unit}[1]{\mbox{$\rm \,#1$}}
\newcommand{\fc}{f_c}
\newcommand{\fm}{f_m}
\newcommand{\U}{U}
\newcommand{\Uc}{U_c}
\newcommand{\thd}{\textnormal{THD}U}
\newcommand{\pst}{P_{st}}
\newcommand{\plt}{P_{lt}}
\newcommand{\fmi}{f_{m_i}}
\newcommand{\Tmi}{T_{m_i}}
\newcommand{\kmi}{k_{m_i}}
\newcommand{\fmic}{f_{m_i c}}
\newcommand{\kmic}{k_{m_i c}}
\newcommand{\dfmi}{\delta f_{m_i}}
\newcommand{\dkmi}{\delta k_{m_i}}
\newcommand{\uin}{u_{IN}\left( t_k \right) }
\newcommand{\umod}{u_{\textnormal{mod}}\left( t_k \right) }
\newcommand{\umodc}{u_{\textnormal{mod} c}\left( t_k \right) }
\newcommand{\umodi}{u_{\textnormal{mod}_i}\left( t_k \right) }
\newcommand{\uacov}{u_{\textnormal{acov}}\left( t_k \right) }
\newcommand{\uxcov}{u_{\textnormal{xcov}}\left( t_k \right) }
\newcommand{\uxcorr}{u_{\textnormal{xcorr}}\left( t_k \right) }
\newcommand{\uacovf}{u_{\textnormal{acov}}\left( f \right) }
\newcommand{\tk}{t_k}
\newcommand{\tkm}{t_{k \textnormal{max}}}
\newcommand{\fs}{f_s}
\newcommand{\N}{N}
\newcommand{\Nj}{N_j}
\newcommand{\sfmjc}{\hat{f}_{m_j c}}
\newcommand{\dfmc}{\Delta f_{m c}}
\newcommand{\fimi}{\varphi_{m_i}}
\newcommand{\fimic}{\varphi_{m_i c}}
\newcommand{\Dd}{\Delta \delta}
\newcommand{\Dk}{\Delta k}
\newcommand{\dmic}{\delta_{m_i c}}
\newcommand{\dmi}{\delta_{m_i}}
\newcommand{\utest}{u_{\textnormal{test}}\left( t_k \right) }
\newcommand{\unoise}{u_{\textnormal{noise}}\left( t_k \right) }
\newcommand{\uc}{u_{c}\left( t_k \right) }
\newcommand{\mc}{m_c}
\newcommand{\Mc}{M_c}
\newcommand{\kU}{k_U}
\newcommand{\depth}{\left( \Delta U_m / U_m \right) }

\begin{document}
	
\IEEEoverridecommandlockouts
\IEEEpubid{\begin{minipage}{\textwidth}\ \\[10pt]
		\centering\footnotesize{\newline \newline \newline \newline \copyright 2023 IEEE. Personal use of this material is permitted. Permission from IEEE must be obtained for all other uses, in any current or future media, including reprinting/republishing this material for advertising or promotional purposes, creating new collective works, for resale or redistribution to servers or lists, or reuse of any copyrighted component of this work in other works. DOI:~10.1109/TIE.2023.3283698}
\end{minipage}}
	
\title{Decomposition by Approximation with Pulse Waves Allowing Further Research on Sources of Voltage Fluctuations}

\author{
	
	Piotr Kuwa{\l{}}ek, \emph{Member, IEEE} \vspace{-2em}
	
	\thanks{
	
		Manuscript received 11 January 2023; revised 14 March 2023 and
		13 April 2023; accepted 26 May 2023. This work was funded by National Science Centre, Poland -- 2021/41/N/ST7/00397. For the purpose of Open Access, the author has applied a CC--BY public copyright licence to any Author Accepted Manuscript ({AAM}) version arising from this submission.
		
		Piotr Kuwa{\l{}}ek is with the Institute of Electrical Engineering and Electronics, Faculty of Control, Robotics and Electrical Engineering, Poznan University of Technology, Poznan, Poland, (e-mail: piotr.kuwalek@put.poznan.pl). 
		
		}
}

\maketitle
	
\begin{abstract}
Voltage fluctuations are common disturbances in power grids. Initially, it is necessary to selectively identify individual sources of voltage fluctuations to take actions to minimize the effects of voltage fluctuations. Selective identification of disturbing loads is possible by using a signal chain consisting of demodulation, decomposition, and assessment of the propagation of component signals. The accuracy of such an approach is closely related to the applied decomposition method. The paper presents a new method for decomposition by approximation with pulse waves. The proposed method allows for an correct identification of selected parameters, that is, the frequency of changes in the operating state of individual sources of voltage fluctuations and the amplitude of voltage changes caused by them. The article presents results from numerical simulation studies and laboratory experimental studies, based on which the estimation errors of the indicated parameters were determined by the proposed decomposition method and other empirical decomposition methods available in the literature. The real states that occur in power grids were recreated in the research. The metrological interpretation of the results obtained from the numerical simulation and experimental research is discussed.
\end{abstract}


\markboth{IEEE TRANSACTIONS ON INDUSTRIAL ELECTRONICS}%
{}

\definecolor{limegreen}{rgb}{0.2, 0.8, 0.2}
\definecolor{forestgreen}{rgb}{0.13, 0.55, 0.13}
\definecolor{greenhtml}{rgb}{0.0, 0.5, 0.0}

\section{Introduction}

\IEEEPARstart{E}{lectrical} energy is one of the basic energy sources used to supply devices. Regardless of power requirements and load characteristics, it is necessary to supply energy with appropriate parameters. This forces power suppliers to distribute energy of a certain quality, allowing for the efficient use of supplied devices. This requirement is defined by the appropriate criteria for parameters determining power quality~\cite{pq_st_50160} (e.g., rms value of voltage~$\U$, fundamental frequency of voltage~$\fc$, total harmonic distortion of voltage~$\thd$, short--term/long--term flicker indicator~$\pst/\plt$). The technical and commercial conditions force the development and use of different methods of power quality evaluation~\cite{tii_pqd_1,decomp,tie_pqd_1,tie_pqd_2}. Common reasons for the deterioration of power quality are voltage fluctuations~\cite{RaportWN}. This phenomenon can cause flicker for different types of light sources (e.g. incandescent lamps, fluorescent lamps, LEDs)~\cite{flicker_ichqp} and can disturb the operation and reduce the life of other loads (e.g. induction motors~\cite{gnacinski_tec_2019}). Depending on the cause, voltage fluctuations occur in power grids from the LV network to the HV network. The causes (sources) of voltage fluctuations are, for example, operations of a specific load (e.g. arc furnace, machine controlling technological process) or their groups (in this case, the resultant frequency~$\fm$ of changes in the operating state of a group of loads can be higher than the power frequency~$\fc$)~\cite{granica150_1,granica150_2,moje_TPD}, changes in the network topology and impedance of the supply circuit. The effects of operations of disturbing loads are changes in the current, which are the direct cause of voltage fluctuations. Voltage fluctuations can also come from other circuits (usually voltage fluctuations propagate from circuit with higher rated voltage to circuit with lower rated voltage). It is worth noting that now also power electronic devices can be a source of voltage fluctuations (even if there is high--frequency switching of the order of~\unit{kHz}), the number of which is significantly increasing (e.g. due to the increase in the number of new installations of renewable energy sources in the power grid)~\cite{elektr1}. Therefore, there is a need to identify the occurring sources of voltage fluctuations, in order to indicate the supply point of individual sources of voltage fluctuations, and then to take actions to minimize the effects of voltage fluctuations.

In the process of identification of sources of voltage fluctuations, information is obtained on the frequency~$\fmi$ of changes in the operating state of individual $i$--th sources of voltage fluctuations in the power grid (a feature dependent on disturbing loads) and on the amplitude~$\kmi$ of voltage changes caused by them (a feature dependent on disturbing loads and their supply circuit). The accurate estimation of indicated parameters~($\fmi$,$\kmi$) supports the diagnostic of voltage fluctuations~\cite{def_wsk,moje_TPD}, and in particular allows for the selective localization of sources of voltage fluctuations in the power grid based on simultaneous series of measurements at particular points in the power grid~\cite{moje_En}. Currently, there are not too many methods in the literature that allow for the achievement of indicated goal (even in a limited way). Unfortunately, the methods available in the literature:
\begin{itemize}
\item limit the identification of disturbing loads to those whose frequency~$\fmi$ of operating state changes is lower than the power frequency~$\fc$~\cite{single_point_1,single_point_5,multi_point_1,lok_wicz}; or
\item can cause incorrect estimation of indicated parameters~($\fmi$,$\kmi$) due to limited possibilities of the decomposition method used (in some cases)~\cite{decomp_limit,lok_wicz}; or
\item have limited diagnostic capabilities (e.g. lack of automation of the identification process, additional expert knowledge required, the need for iterative procedures in which the dominant disturbing load is identified and the effect of changes in its operating state is eliminated)~\cite{single_point_1,multi_point_3,multi_point_2,lok_wicz}.
\end{itemize}

The paper presents a new proprietary method of decomposition by approximation with pulse waves (DAPW), the description of which is presented in Section~\ref{sec2}. The presented method allows for the accurate identification of selected parameters, i.e. the frequency~$\fmi$ of changes in the operating state of individual sources of voltage fluctuations and the amplitude~$\kmi$ of voltage changes caused by them, which allows for automatic selective identification and localization of many sources of voltage fluctuations based on simultaneous series of measurements at particular points of the power grid (without additional expert knowledge). Section~~\ref{sec3} presents the results of numerical simulation studies and laboratory experimental studies, on the basis of which were determined the estimation errors~($\dfmi$,$\dkmi$) of indicated parameters~($\fmic$,$\kmic$) estimated by the proposed decomposition method and by other empirical decomposition methods available in the literature, which have not yet been used for this purpose other than Enhanced Empirical Wavelet Transform (EEWT). The presented research results show that the proposed approach is characterized by the smallest estimation errors~($\dfmi$,$\dkmi$) of the indicated parameters, thanks to which the use of the proposed method in the process of selective identification of voltage fluctuations sources based on a demodulation with a carrier signal estimation, allows for the accurate estimation of selected features of disturbing loads in the power grid (including power electronic devices).

\section{Proposed Approach}\label{sec2}

\begin{figure}[ht]\centering
	\includegraphics[width=.5\columnwidth]{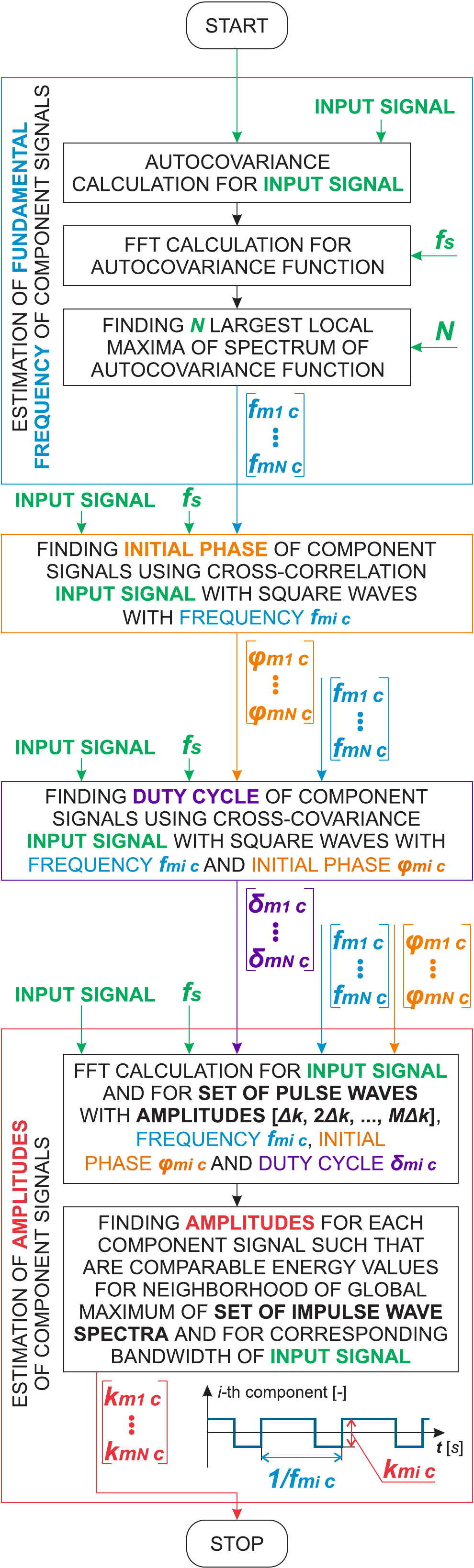}
	\caption{The simplified block diagram of the proposed approach}\label{Fig::algorithm}
\end{figure}

A simplified diagram of the proposed proprietary decomposition approach using approximation by pulse waves is shown in Fig.~\ref{Fig::algorithm} (the colors of graphic representation for the individual steps of the proposed approach are maintained). The idea of the proposed approach is that the amplitude modulating signal of supply voltage in the power grid is approximated by a series of pulse waves. The selection of basis functions in the form of pulse waves results from the fact that probably most present sources of voltage fluctuations cause step (rectangular) voltage changes (chaotic sources causing irregular voltage changes currently have mostly separate supply circuits or their effects are already minimized by appropriate equipment). In the proposed decomposition process, information about the actual shape of voltage changes caused by individual disturbing loads is lost. Nevertheless, the assumption of the pulse waves as the basis functions allows for the accurate extraction of the two most important parameters~($\fmi$,$\kmi$) used in the process of indication of the supply points of individual sources of voltage fluctuations. The amplitude~$\kmic$ of the estimated $i$--th pulse wave corresponds to the maximum voltage change~$\kmi$ caused by the operation of the $i$--th disturbing load, and the fundamental frequency~$\fmic$ of the estimated $i$--th pulse wave corresponds to the average value of frequency~$\fmi$ of the quasi--periodic changes in the operating state of $i$--th disturbing load.

\begin{figure}[ht]\centering
	\centering
	\begin{subfigure}{.5\textwidth}
		\centering
		\includegraphics[width=.6\columnwidth]{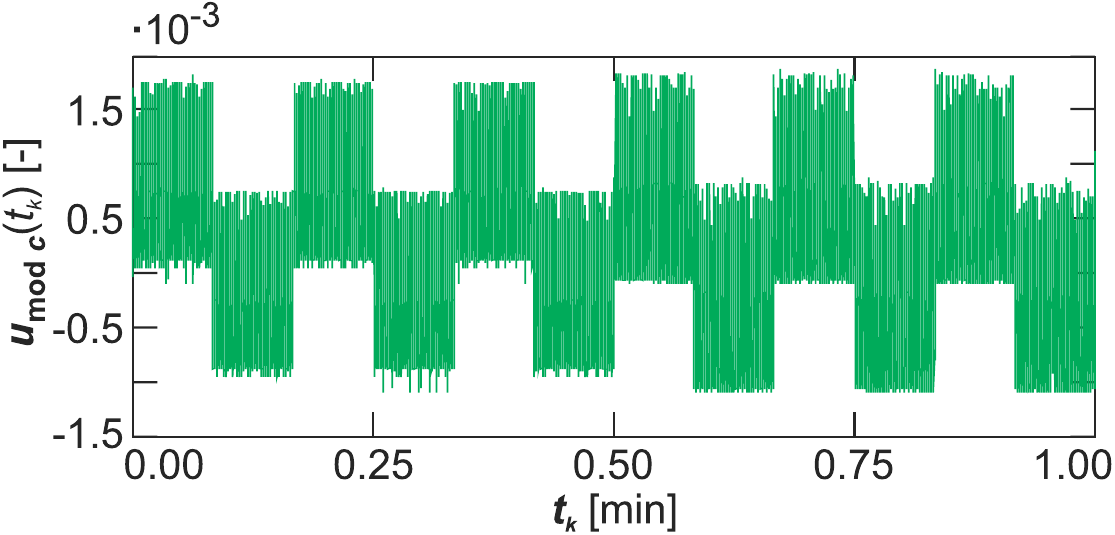}
		\caption{recreated amplitude modulating signal~$\umodc$}
	\end{subfigure}
	\begin{subfigure}{.5\textwidth}
		\centering
		\includegraphics[width=.6\columnwidth]{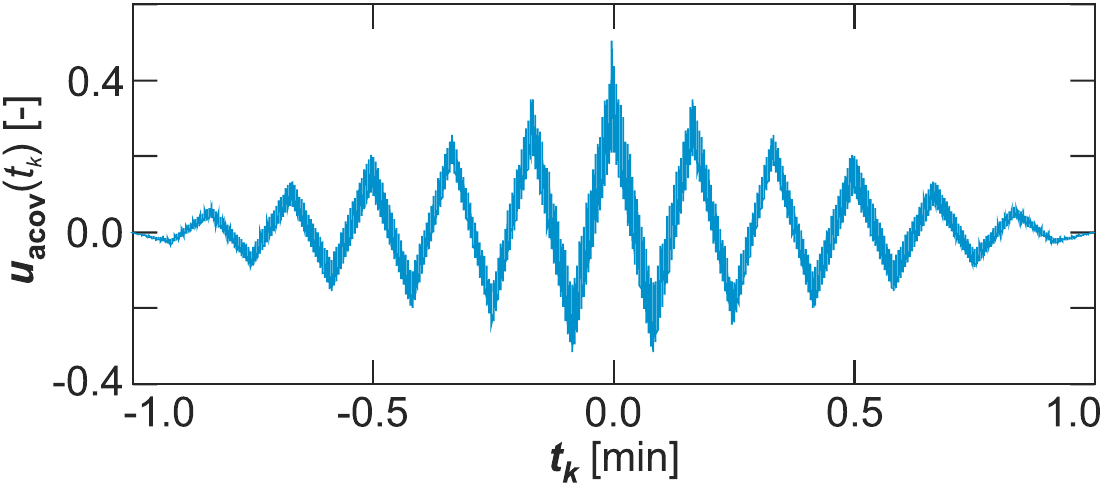}
		\caption{autocovariance function~$\uacov$ of the recreated signal~$\umodc$}
	\end{subfigure}
	\begin{subfigure}{.5\textwidth}
		\centering
		\includegraphics[width=.6\columnwidth]{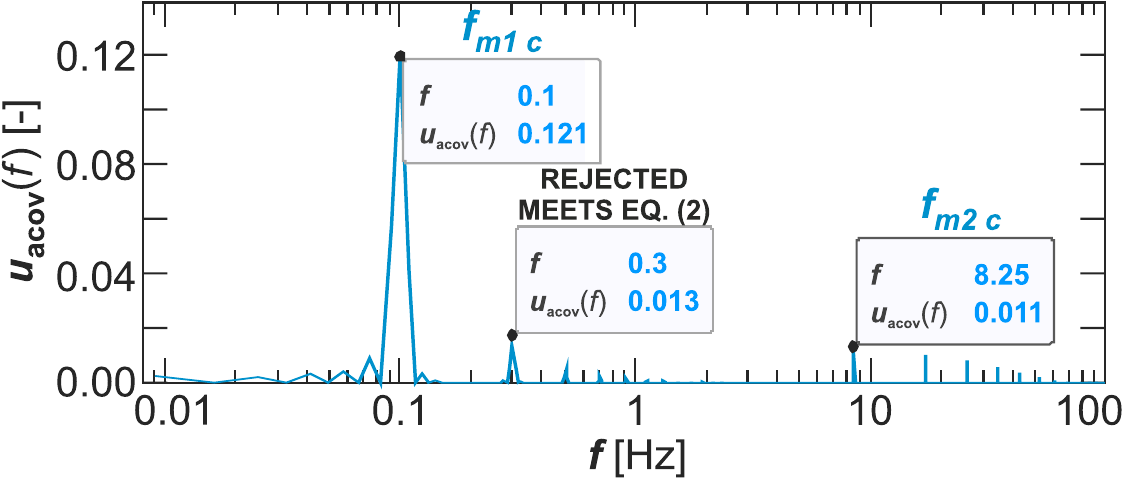}
		\caption{spectrum~$\uacovf$ of the determined function~$\uacov$ -- selection of appropriate $\fmic$~values}	
	\end{subfigure}
	\caption{The example of procedure for estimation of frequencies~$\fmic$ of basis pulse waves, where $\N$=2}\label{Fig::stage1_f}
\end{figure}

In the first stage of the proposed approach, the fundamental frequencies~$\fmic$ for particular $i$--th pulse waves are estimated. For this purpose, for the input signal~$\uin$ (in terms of diagnostics of voltage fluctuations, the input signal~$\uin$ is the estimated amplitude modulating signal~$\umod$, obtained using the demodulation with estimation of carrier signal), the autocovariance function~$\uacov$ for individual samples~$\tk$ is determined (sample interval $\Delta t = 1 / \fs$, where~$\fs$ is sampling rate) according to the relation~\cite{sig_process}:

{\small
\begin{align}\label{acov}
	\uacov = \frac{1}{L}\sum_{i=1}^{L-k} \Bigg\{ & \left( u_{IN}\left( t_i \right) - \frac{1}{L} \sum_{j=1}^{L} u_{IN}\left( t_j \right)   \right) \\ \nonumber & \cdot \left( u_{IN}\left( t_{i+k} \right) - \frac{1}{L} \sum_{j=1}^{L} u_{IN}\left( t_j \right)   \right) \Bigg\} \,,
\end{align}
}

\noindent
where~$L$ is the number of samples in the measurement window.
\noindent
For the obtained autocovariance function, the signal spectrum is determined using fast Fourier transform (FFT) (if the number of samples does not allow for the implementation of the Cooley--Tukey algorithm, discrete Fourier transform (DFT) algorithm is used instead of FFT). In order to estimate the fundamental frequencies~$\fmi$ for $\N$~pulse waves, the local maxima~$\sfmjc$ of the autocovariance function spectrum are determined and $\N$~maxima with highest values are selected, except that those maxima are rejected that meet the relation:
\begin{align}\label{reject_rel}
	\forall_{j,k \in 1,...,N_j:\left( j \neq k\right) } \hat{f}_{m_j c} \cong n \hat{f}_{m_k c} \,,
\end{align}
where: $n \in \left\lbrace 2,3,4,5 \right\rbrace $, $\Nj$ is the number of determined local maxima of the spectrum of autocovariance function~$\uacov$. The approximately equal in dependence~(\ref{reject_rel}) is assumed to be equal with the resolution~$\dfmc$:
\begin{align}
	\dfmc = \frac{10 \fs}{L} .
\end{align}
\noindent
An example of procedure in the first stage is shown in Fig.~\ref{Fig::stage1_f}.

\begin{figure}[!]\centering
	\centering
	\begin{subfigure}{0.5\textwidth}
		\centering
		\includegraphics[width=.6\columnwidth]{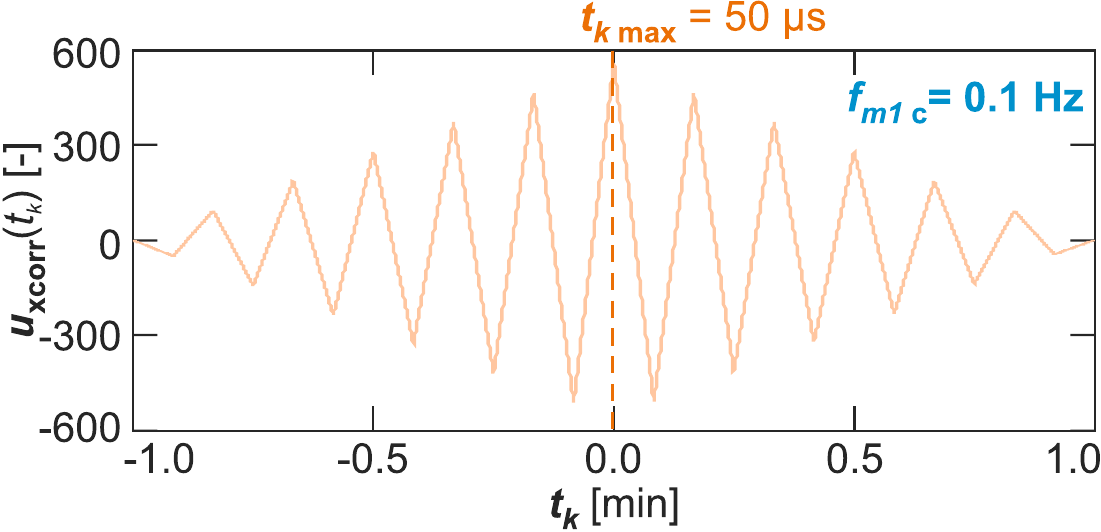}
		\caption{cross--correlation function~$\uxcorr$ for input signal~$\uin$~=~$\umodc$ and square wave with $f_{m1c}$=0.1~\unit{Hz}}
	\end{subfigure}
	\begin{subfigure}{0.5\textwidth}
		\centering
		\includegraphics[width=.6\columnwidth]{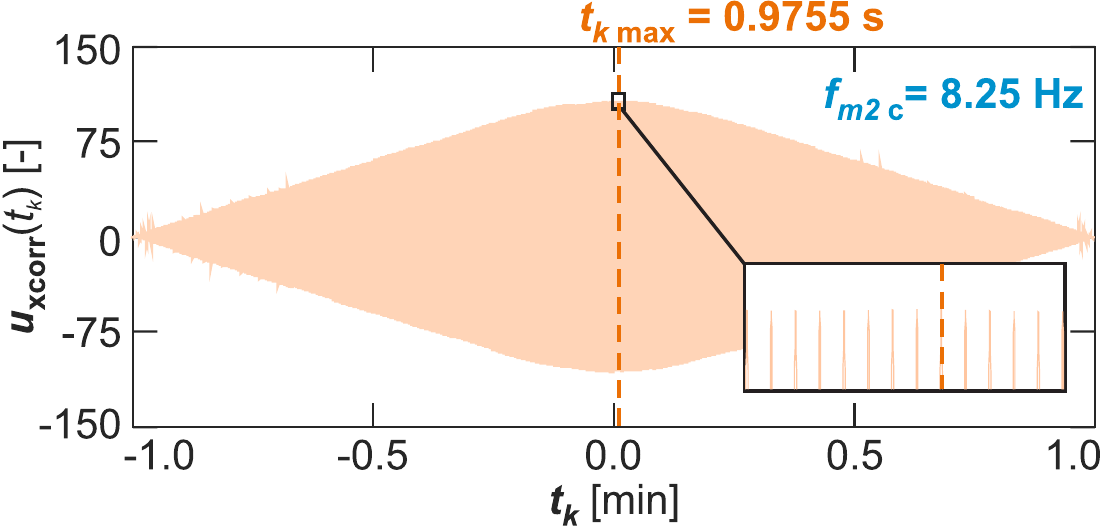}
		\caption{cross--correlation function~$\uxcorr$ for input signal~$\uin$~=~$\umodc$ and square wave with $f_{m2c}$=8.25~\unit{Hz}}
	\end{subfigure}
\caption{The example of procedure for estimation of initial phases~$\fimic$ of basis pulse waves, where $\N$=2}\label{Fig::stage2_fi}
\end{figure}

In the second stage, initial phases~$\fimic$ are estimated for individual component signals with frequencies~$\fmic$ determined in the first stage. In order to determine the initial phase~$\fimic$ for the $i$--th pulse wave with fundamental frequency~$\fmic$, the cross--correlation function~$\uxcorr$ between the input signal~$\uin$~=~$\umodc$ and the normalized square wave with frequency~$\fmic$ is determined for individual samples~$\tk$ according to the dependence~\cite{sig_process}:
{\small
\begin{align}
	\uxcorr = \frac{1}{L} \sum_{i=1}^{L-k}\left\lbrace u_{IN}\left( t_i \right) \cdot \text{sign}\left( \sin \left( 2 \pi \fmic t_{i+k} \right) \right)  \right\rbrace .
\end{align}
}
\noindent
In the next step, based on the determined cross--correlation function~$\uxcorr$, the global maximum~$\tkm$ is determined. The value of global maximum~$\tkm$ is the delay time expressed in seconds. In order to calculate the initial phase~$\fimic$ in radians based on the delay time~$\tkm$, the following relationship should be used:
\begin{align}
	\fimic = \mod \left( 2 \pi \tkm \fmic , 2 \pi \right) \left[ \text{rad} \right] . 
\end{align}
\noindent
An example of procedure in the second stage is shown in Fig.~\ref{Fig::stage2_fi}.

In the third stage, the duty cycles~$\dmic$ are estimated for individual pulse waves. For this purpose, for each $i$--th basis pulse waves, a set of pulse waves is created with an amplitude equal to half of the peak--to--peak value of the input signal~$\uin$~=~$\umodc$, with the fundamental frequency~$\fmic$, with the initial phase~$\fimic$, and with the duty cycle~$\left[ \Dd, 2 \Dd, ... , 1 - \Dd  \right] $, where $\Dd$~is the adopted resolution with which the unknown value of the duty cycle of a particular basis pulse wave is estimated. The resolution $\Dd$=0.01 is arbitrarily assumed in the research. As the estimated value of the duty cycle~$\dmic$ of $i$--th basis pulse wave, the value of~$n \Dd$ $\left( n \in N \land n : n \Dd \in \left( 0 ; 1\right) \right) $ is assumed, for which the largest value of the global maxima is obtained for individual cross--covariance functions~$\uxcov$ between input signal~$\uin$~=~$\umodc$ and a properly created set of pulse waves. To calculate the cross--covariance between any two signals $u_1 \left(  t \right) $ and $u_2 \left(  t \right) $ for individual samples~$\tk$ is used the relationship~\cite{sig_process}:

{\small
\begin{align}\label{xcov}
	\uxcov = \frac{1}{L}\sum_{i=1}^{L-k} \Bigg\{ & \left( u_{1}\left( t_i \right) - \frac{1}{L} \sum_{j=1}^{L} u_{1}\left( t_j \right)   \right) \\ \nonumber & \cdot \left( u_{2}\left( t_{i+k} \right) - \frac{1}{L} \sum_{j=1}^{L} u_{2}\left( t_j \right)   \right) \Bigg\} .
\end{align}
}

\begin{figure}[htb]\centering
	\centering
	\includegraphics[width=.6\columnwidth]{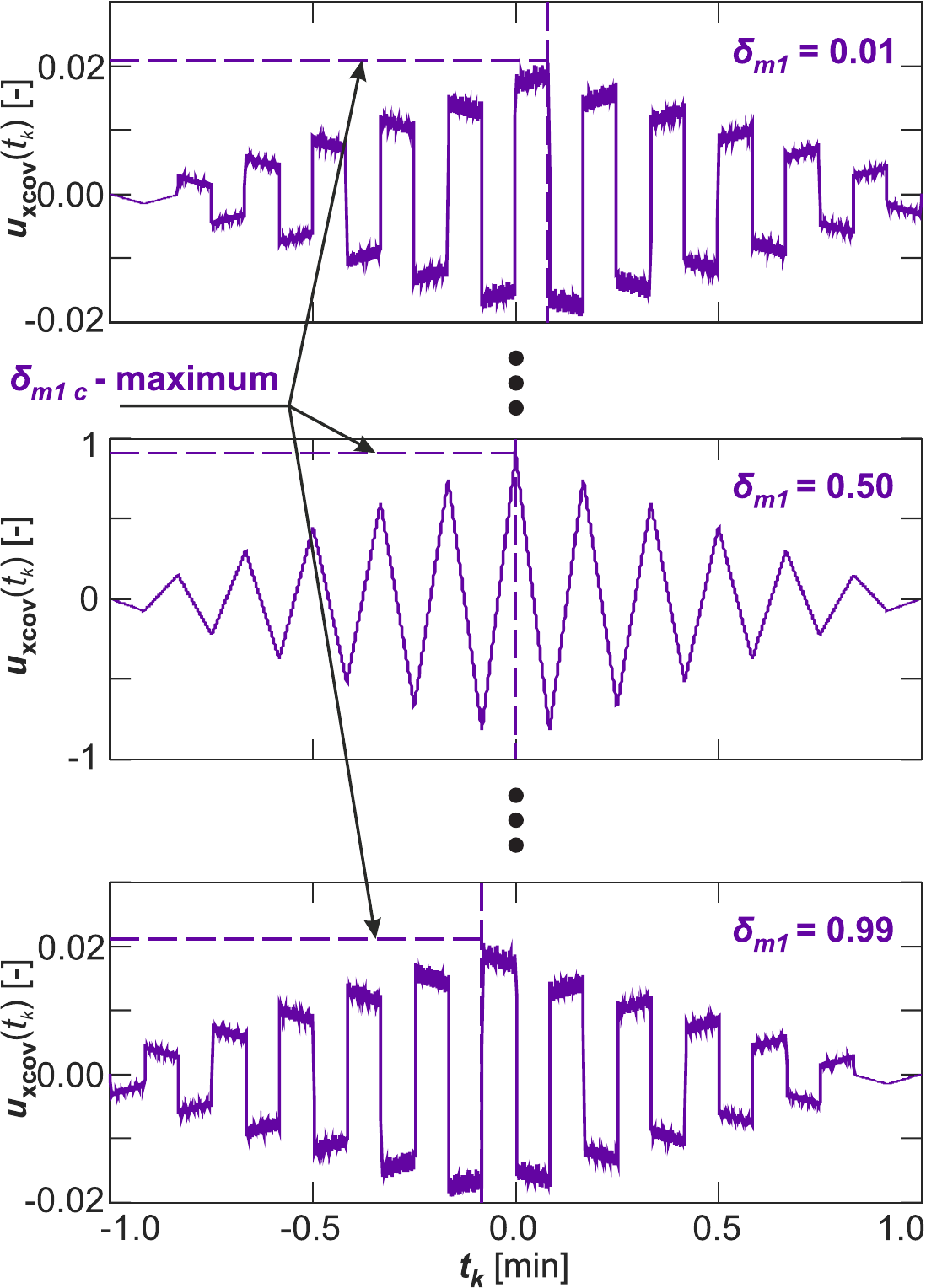}
	\caption{The example of procedure for estimation of duty cycle~$\delta_{m_1 c}$ of basis pulse wave, where $\N$=2, $f_{m_1 c}$=0.1\unit{Hz}, $\varphi_{m_1 c}$=0~\unit{rad}}\label{Fig::stage3_d}
\end{figure}

\noindent
An example of procedure in the third stage is shown in Fig.~\ref{Fig::stage3_d}.

\begin{figure}[htb]\centering
	\centering
	\includegraphics[width=.6\columnwidth]{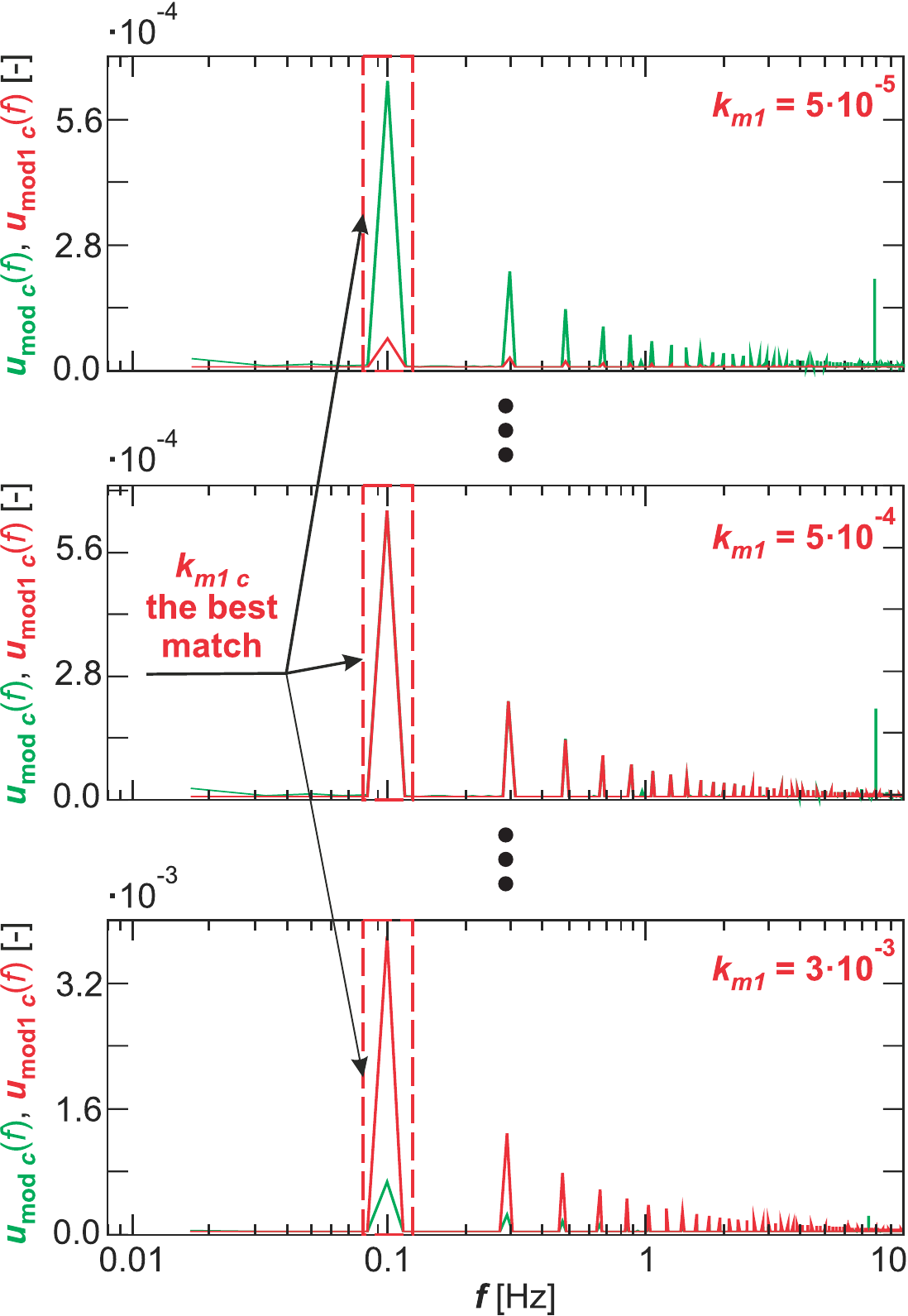}
	\caption{The example of procedure for estimation of amplitude~$k_{m_1 c}$ of basis pulse wave, where $\N$=2, $f_{m_1 c}$=0.1\unit{Hz}, $\varphi_{m_1 c}$=0~\unit{rad}, $\delta_{m_1 c}$=0.50}\label{Fig::stage4_k}
\end{figure}

In the last stage, the amplitudes~$\kmic$ are estimated for individual $i$--th basis pulse waves. For this purpose, the spectra are determined using the FFT algorithm (if the number of samples does not allow for the implementation of Cooley--Tukey algorithm, DFT is used instead of FFT) for the input signal~$\uin$~=~$\umodc$ and for a set of pulse waves with the fundamental frequency~$\fmic$, initial phase~$\fimic$, duty cycle~$\dmic$, and with amplitude $\left[ \Dk, 2 \Dk, ... , R \Dk  \right] $, where $\Dk$~is the adopted resolution with which the unknown value of the amplitude of a particular basis pulse wave is estimated, and $R$ is such a natural number that the value of $R \Dk$ is equal to the peak--to--peak value of the input signal~$\uin$~=~$\umodc$. As an estimator of the unknown amplitude~$\kmic$ for the $i$--th basis pulse wave is assumed the value for which the energy value for the neighborhood of the global maximum of the spectrum from the set of pulse waves and the energy value for the corresponding bandwidth of the input signal~$\uin$~=~$\umodc$ is comparable. An example of procedure in the fourth stage is shown in Fig.~\ref{Fig::stage4_k}.

\section{Research Results and Discussion}\label{sec3}

\subsection{Test Signal}

For the verification of the proposed approach, as the test signal is selected signal described by the relationship:
\begin{align}\label{mod_eq}
	\utest = \uc \cdot \left( 1 + \umod \right) .
\end{align}

\begin{figure}[htb]
	\centering
	\includegraphics[width=0.6\columnwidth]{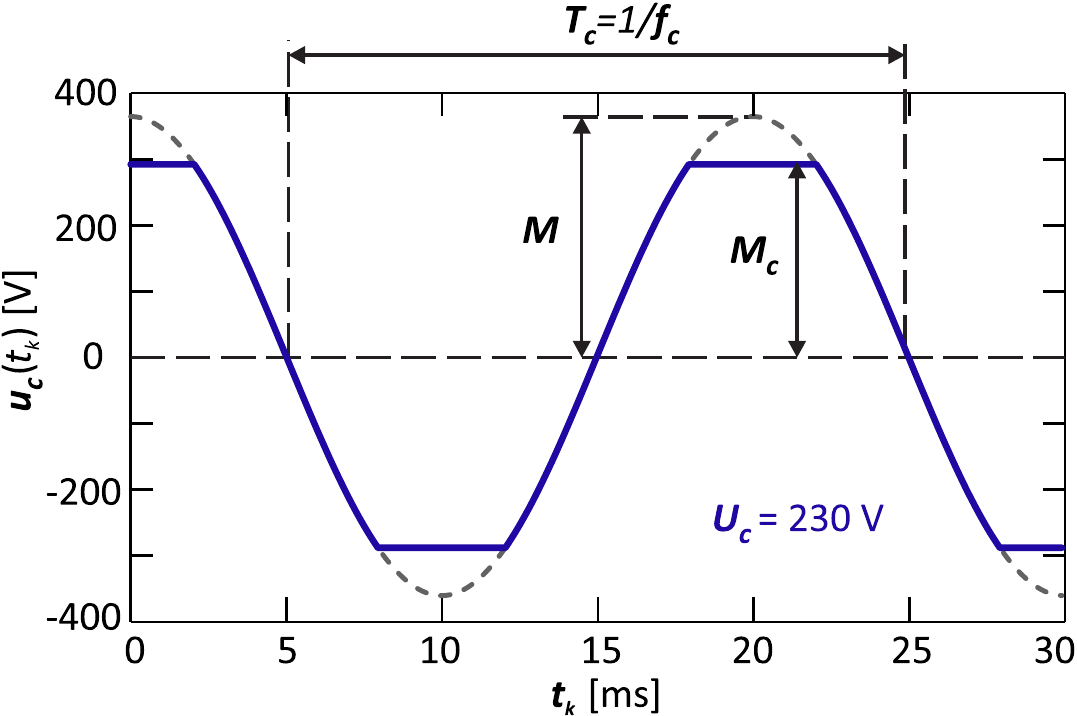}
	\caption{The exemplary waveform of ``clipped cosine" for $\mc=0.8$ and $\kU \approx 361.6$}
	\label{fig:ccos}
\end{figure}

\noindent
The relationship described by~(\ref*{mod_eq}) defines amplitude modulation (AM) without attenuated carrier wave and is a properly representation of voltage fluctuations that occur in the real stiff power grid~\cite{def_wsk,moje_TPD,moje_En}, i.e., where the power frequency has negligible deviations (otherwise it is necessary to recreate voltage fluctuations as amplitude--phase/frequency modulation (AM--PM/FM)~\cite{bien_am_fm}). The carrier signal~$\uc$ in~(\ref{mod_eq}) describes the supply voltage before voltage fluctuations occur. In order to recreate the states occurring in the real power grid, the ``clipped cosine" type signal is adopted as a carrier signal~$\uc$, which is associated with the effect of input stages of switching power supplies. Therefore, the carrier signal~$\uc$ is given by:

{\footnotesize
\begin{align}\label{ccos}
	\uc =\left\{ \begin{matrix}
		{\kU}{\mc} & \text{if} & \cos \left( 2\pi {\fc} \tk \right)>{\mc}  \\
		{\kU}\cos \left( 2\pi {\fc} \tk \right) & \text{if} & -{\mc}\ge \cos \left( 2\pi {\fc} \tk \right)\le {\mc}  \\
		-{\kU}{\mc} & \text{if} & \cos \left( 2\pi {\fc} \tk \right)<-{\mc}  \\
	\end{matrix} \right.
	,
\end{align}
}

\noindent
where $\fc$~is the carrier frequency ($\fc=50$\unit{Hz} was adopted in the research); $\kU$ is a scaling value which, for a given value of $\mc$, allows conversion the rms value of test signal to the rated value ($\Uc$=230\unit{V} was adopted in the research, i.e., the rated rms value in the {LV} network in Europe was adopted in the research); $\mc$ defines the clipping level and is given by:
\begin{align}\label{mc}
	\mc =\frac{\Mc}{M}\le 1,
\end{align}
where: $\Mc$ and $M$ are amplitudes after and before clipping, respectively. The value of $\mc$=0.8 is adopted in the research, which corresponds to the occurrence of voltage distortion at the level of the limit of acceptable distortion in the LV~network~\cite{pq_st_50160}. An example of the adopted carrier signal~$\uc$ is shown in Fig.~\ref{fig:ccos}.

\begin{figure}[htb]
	\centering
	\includegraphics[width=0.6\columnwidth]{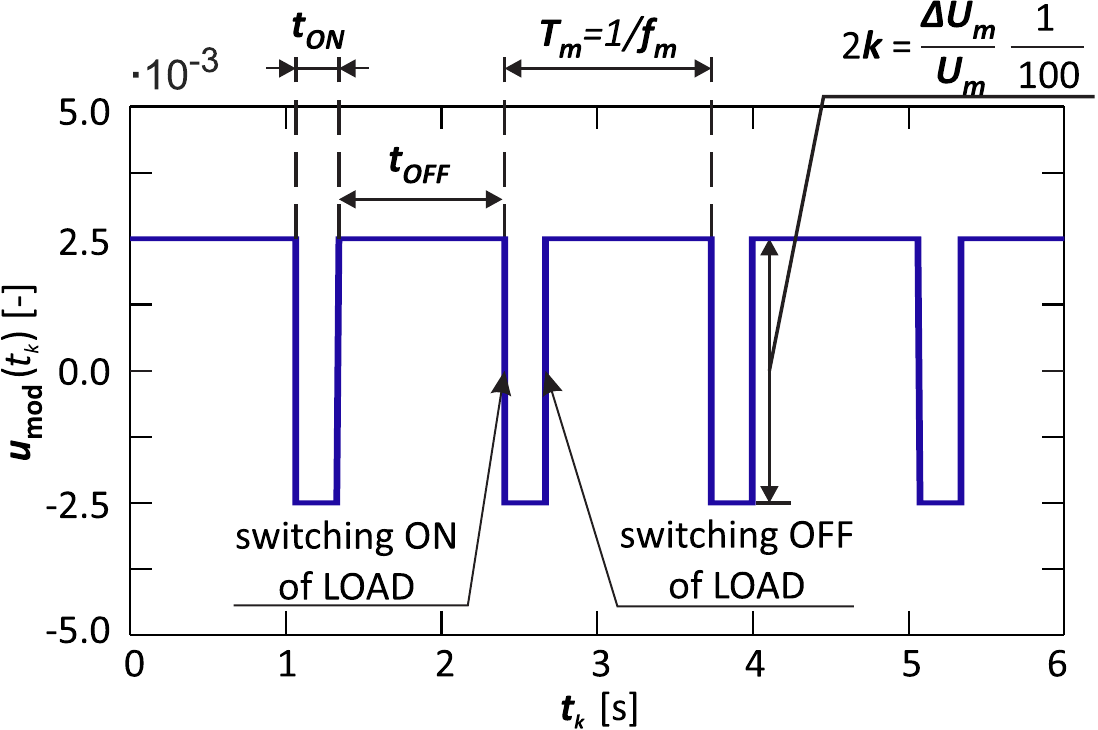}
	\caption{The example of adopted component of modulating signal}
	\label{fig:umodi}
\end{figure}

\begin{figure}[htb]
	\centering
	\includegraphics[width=0.6\columnwidth]{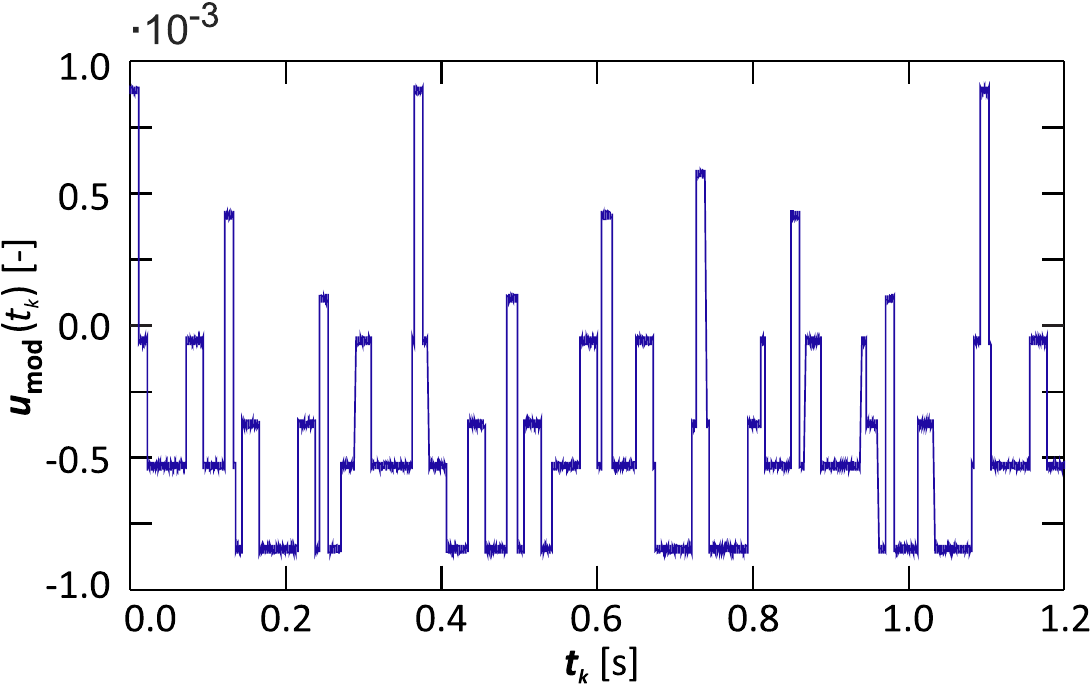}
	\caption{The example of adopted resultant modulating signal associated with the operation of three dominant sources of voltage fluctuations}
	\label{fig:umod}
\end{figure}

The modulating signal~$\umod$ in~(\ref{mod_eq}) is a signal associated with the resultant operation of $\N$~disturbing loads. In the research, the modulating signal~$\umod$ was adopted as the sum of $\N$ asymmetric rectangular signals~$\umodi$ (a signal associated with step voltage changes, which are probably the most common in the power grid) with a duty cycle~$\dmi$:
\begin{align}
	\dmi = \frac{t_{{ON}_i}}{\Tmi},
\end{align}
associated with individual $i$--th disturbing loads, and noise~$\unoise$ with uniform distribution and standard deviation equal to~$10^{-5}$. Therefore, the modulating signal~$\umod$ is given by:
\begin{align}
	\umod = \sum_{i=1}^{N}\umodi + \unoise ,
\end{align}
\begin{align}
	\umodi =\left\{ \begin{matrix}
	\kmi & \text{if} & l\Tmi < \tk < l\Tmi + t_{ON}  \\
	-\kmi & \text{if} & l\Tmi + t_{ON} \le \tk \le \left( l + 1 \right) \Tmi \\
\end{matrix} \right.
,
\end{align} 
\noindent
where $l \in \mathbb{Z}$. The exemplary component~$\umodi$ of the modulating signal~$\umod$ is shown in Fig.~\ref{fig:umodi}. Fig.~\ref{fig:umod} shows an exemplary resultant modulating signal~$\umod$, assuming the occurrence of three disturbing loads ($\N$=3).

\subsection{Numerical Simulation Studies}

In numerical simulation studies, a set of 9000 test signals was generated in accordance with the relationship defined by~(\ref{mod_eq}), with the following parameters:
\begin{itemize}
	\item the amplitude~$\kmi$ of the $i$--th component~$\umodi$ of modulating signal~$\umod$ selected randomly from the range $\left[ 5 \cdot 10^{-4} ; 2.5 \ 10^{-2} \right] $, which corresponds to the amplitude modulation depth~$\depth$ in the set $\left[ 0.1 ; 5 \right] \% $;
	\item the fundamental frequency~$\fmi$ of the $i$--th component~$\umodi$ of modulating signal~$\umod$ selected randomly from the set $\left[ 0.1 ; 150 \right]$\unit{Hz} (150\unit{Hz} corresponds to 3$\fc$ for $\fc$=50\unit{Hz}~\cite{granica150_1,granica150_2});
	\item the duty cycle~$\dmi$ of the $i$--th component~$\umodi$ of modulating signal~$\umod$ selected randomly from the set $\left\lbrace 0.1,0.2,...,0.9\right\rbrace $;
	\item an initial phase~$\fimi$ of the $i$--th component~$\umodi$ of modulating signal~$\umod$ selected randomly from the set $\left[ 0 ; 2\pi \right) $\unit{rad};
	\item the number~$\N$ of components~$\umodi$ of modulating signal~$\umod$ belonging to the set $\left\lbrace 2 , 3 , 4\right\rbrace $.
\end{itemize}

\noindent
In the process of generation of test signals, cases in which fundamental frequencies~$\fmi$ of individual component signals are equal were avoided. The sampling rate~$\fs$ of individual test signals is 20\unit{kSa/s}, and the time duration of measurement window is 1\unit{min}. According to~\cite{moje_TPD}, this is an enough time for correct extraction of slow voltage changes of 0.1\unit{Hz}. Numerical simulation studies were carried out in the MATLAB software on a computer with an Intel Core i5-1035G1 processor with a clock rate of 3.6\unit{GHz} and 16\unit{GB} of RAM.

\subsection{Laboratory Experimental Studies}

In laboratory experimental studies, a set of 4500 test signals was generated in accordance with the relationship defined by~(\ref{mod_eq}), with the following parameters:
\begin{itemize}
	\item the amplitude~$\kmi$ of the $i$--th component~$\umodi$ of modulating signal~$\umod$ selected randomly from the range $\left[ 5 \cdot 10^{-4} ; 2.5 \cdot 10^{-2} \right] $, which corresponds to the amplitude modulation depth~$\depth$ in the set $\left[ 0.1 ; 5 \right] \% $;
	\item the fundamental frequency~$\fmi$ of the $i$--th component~$\umodi$ of modulating signal~$\umod$ selected randomly from the set $\left[ 0.1 ; 150 \right]$\unit{Hz} (150\unit{Hz} corresponds to 3$\fc$ for $\fc$=50\unit{Hz}~\cite{granica150_1,granica150_2});
	\item the duty cycle~$\dmi$ of the $i$--th component~$\umodi$ of modulating signal~$\umod$ equal to 0.5;
	\item an initial phase~$\fimi$ of the $i$--th component~$\umodi$ of modulating signal~$\umod$ equal to 0\unit{rad};
	\item the number~$\N$ of components~$\umodi$ of modulating signal~$\umod$ belonging to the set $\left\lbrace 2 , 3 , 4\right\rbrace $.
\end{itemize}

\noindent
Fundamental frequencies~$\fmi$ of the $i$--th components~$\umodi$ of modulating signal~$\umod$ were selected randomly, but the following conditions were maintained:
\begin{itemize}
	\item the fundamental frequency~$\fmi$ expressed in\unit{cpm} is a natural number;
	\item least common multiple for~$\fmi$ where $i$=1,..,$\N$ and for $\fc$ expressed in\unit{cpm} is different from 1.
\end{itemize}

\begin{figure}[htb]
	\centering
	\includegraphics[width=0.6\columnwidth]{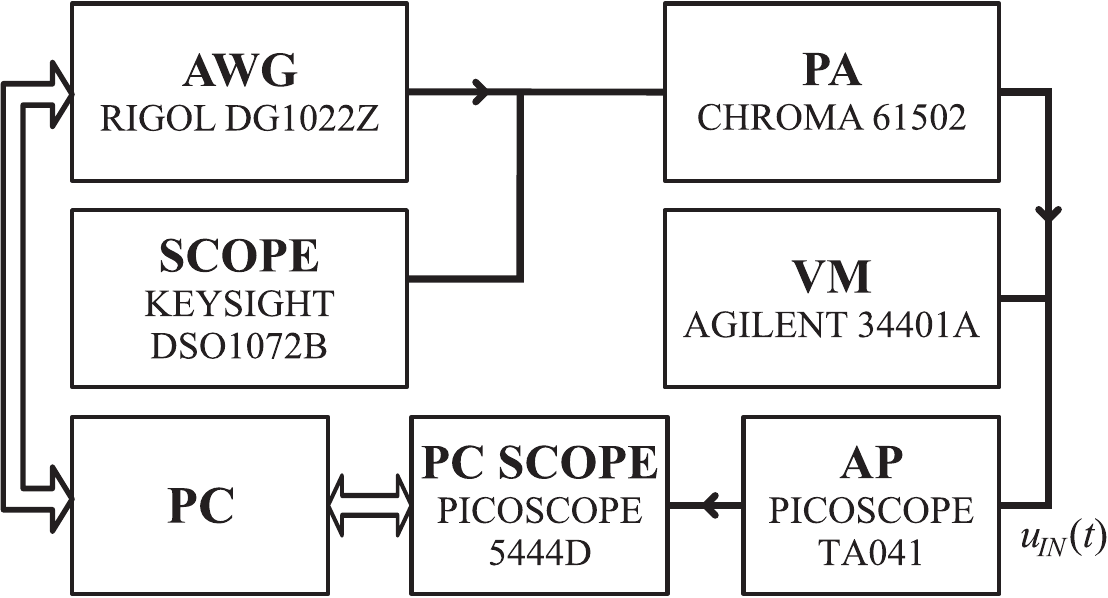}
	\caption{The block diagram of the laboratory setup, where: AWG is an arbitrary waveform generator, PA is a power amplifier, VM is a voltmeter, AP is an active differential probe}
	\label{fig:diagram_lab}
\end{figure}

\begin{figure}[htb]
	\centering
	\includegraphics[width=.7\columnwidth]{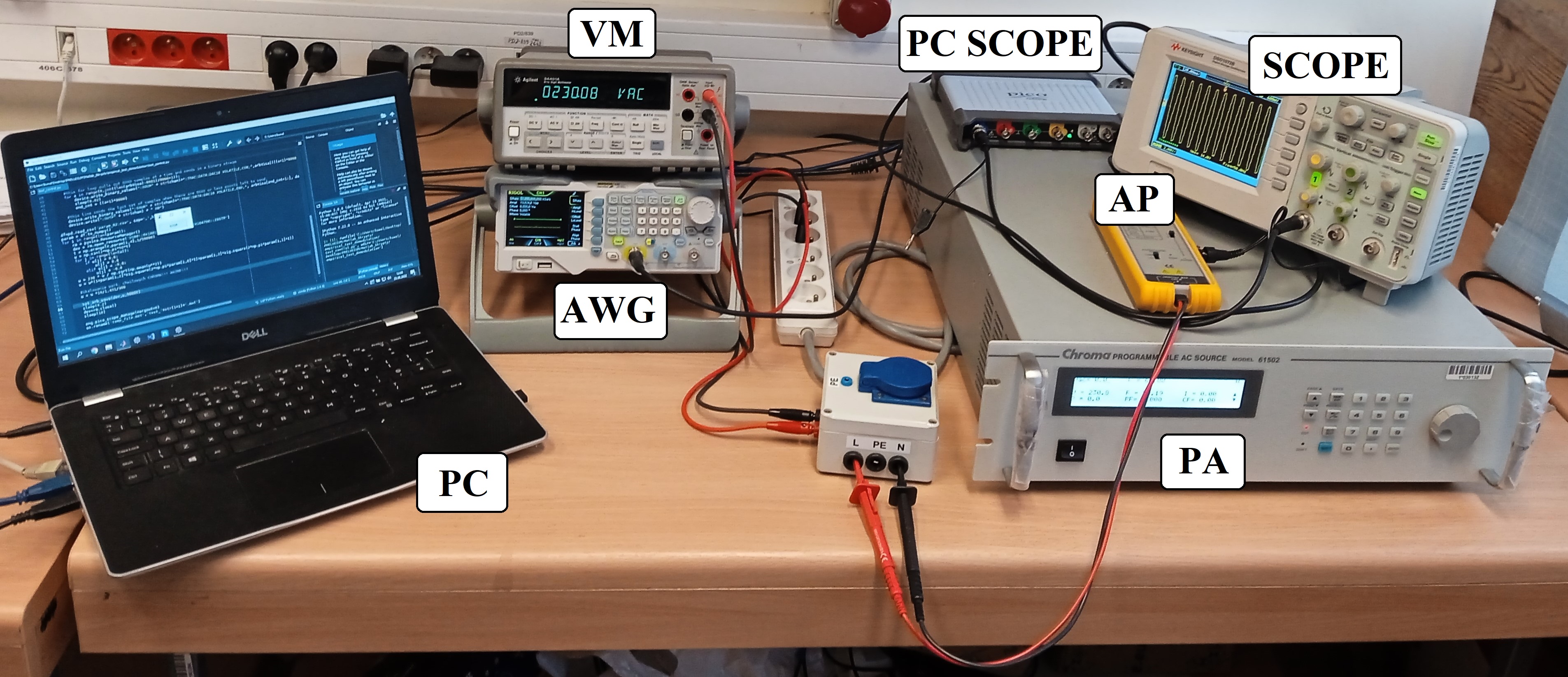}
	\caption{The photo of the laboratory setup}
	\label{fig:photo_lab}
\end{figure}

\noindent
In addition, the standard deviation of noise~$\unoise$ was adopted to be zero. Such conditions allowed for the correct generation of test signals on the laboratory setup, the block diagram of which is shown in Fig.~\ref{fig:diagram_lab}. The photo of the laboratory setup is shown in Fig.~\ref{fig:photo_lab}. In the process of generation of test signals, cases in which the fundamental frequencies~$\fmi$ of individual component signals~$\umodi$ are equal were avoided. The sampling rate~$\fs$ of individual test signals is 50\unit{kSa/s}, and the time duration is 2\unit{min} (in the research, the first minute of the recorded signal was rejected due to the possible occurrence of transient states related to the imperfection of the laboratory equipment used).

\subsection{Research Results}

In the first step, each test signal from the generated set was demodulated using the amplitude demodulation with a carrier signal estimation~\cite{moje_TIE}. Then, the individual recreated modulating signals were decomposed by: the proposed decomposition method described in Section~\ref{sec2}, i.e., decomposition by approximation with pulse waves (DAPW), empirical wavelet transform (EWT)~\cite{ewt}, enhanced empirical wavelet transform (EEWT)~\cite{eewt}, empirical mode decomposition (EMD)~\cite{emd}, variational mode decomposition (VMD)~\cite{vmd}, multidimensional variational mode decomposition (MVMD)~\cite{mvmd}, empirical Fourier decomposition (EFD)~\cite{efd}. For each component signal obtained from the decomposition process, the fundamental frequency~$\fmic$ of $i$--th component signal and its amplitude~$\kmic$ in the considered 1\unit{min} measurement window were determined using particular methods of decomposition. 

\noindent 
To determine the fundamental frequency~$\fmic$ of $i$--th component signal obtained from decomposition methods other than DAPW, the autocorrelation function was used with a~window function that allows determining the~fundamental frequency of any function~\cite{f_measure}. In this research, the~window function was assumed to be a~moving average function. The~selection of the~proposed method for estimation of the~fundamental frequency is due to the~generation of a~quasi--periodic signal as a~result of the~decomposition process. In addition, some types of decomposition showed numerous undulations around the~zero value, which prevented the~use of basic algorithm of frequency detection, i.e., the~zero crossing detector. 

\noindent
The median value of the local maxima of the absolute value of component signal~$\umodi$ (obtained from decomposition methods other than DAPW) after subtracted its average value was used to determine the amplitude~$\kmic$.

On the basis of estimated values of~$\fmic$ and~$\kmic$, the values of relative estimation errors of indicated parameters were determined according to the relationship:
\begin{align}
	\dfmi = \frac{\left| \fmic - \fmi \right| }{\fmi},
\end{align}
\begin{align}
	\dkmi = \frac{\left| \kmic - \kmi \right| }{\kmi}.
\end{align}
\noindent
For the analysis of accuracy of decomposition process, only errors in the estimation of amplitudes~$\kmic$ and frequencies~$\fmic$ of component signals are focused, because these parameters allow for selective identification of individual disturbing loads in the power grid~\cite{moje_En,moje_TPD}. It is worth noting that the determined relative errors include error caused by the decomposition method used and error caused by the demodulation~\cite{moje_TIE} used in the process of recreation of amplitude modulating signal. This approach is important from the point of view of the process of selective identification of sources of voltage fluctuations in the power grid~\cite{moje_En,moje_TPD}, because it is necessary to extract the $\fmic$~and~$\kmic$~values by the proposed signal chain ``demodulation -- decomposition -- statistic assessment of propagation of voltage fluctuations" in~\cite{moje_En}, with the smallest errors~$\dfmi$~and~$\dkmi$.

\begin{figure}[htb]\centering
	\centering
	\begin{subfigure}{0.5\textwidth}
		\centering
		\includegraphics[width=.6\columnwidth]{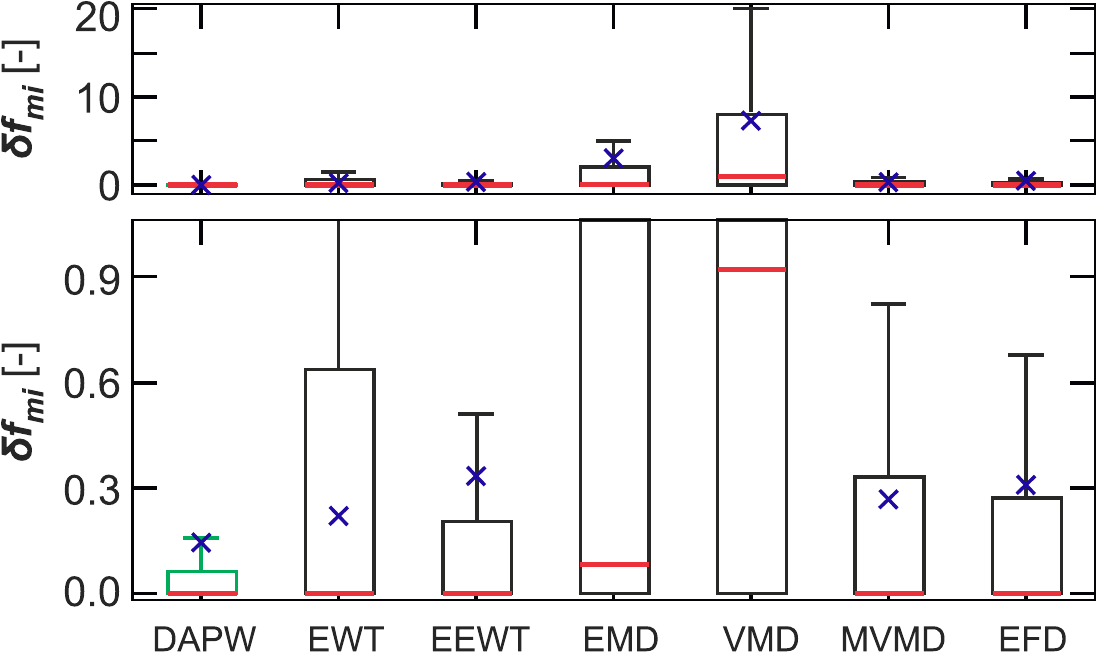}
		\caption{$\N$=2}
	\end{subfigure}
	\begin{subfigure}{0.5\textwidth}
		\centering
		\includegraphics[width=.6\columnwidth]{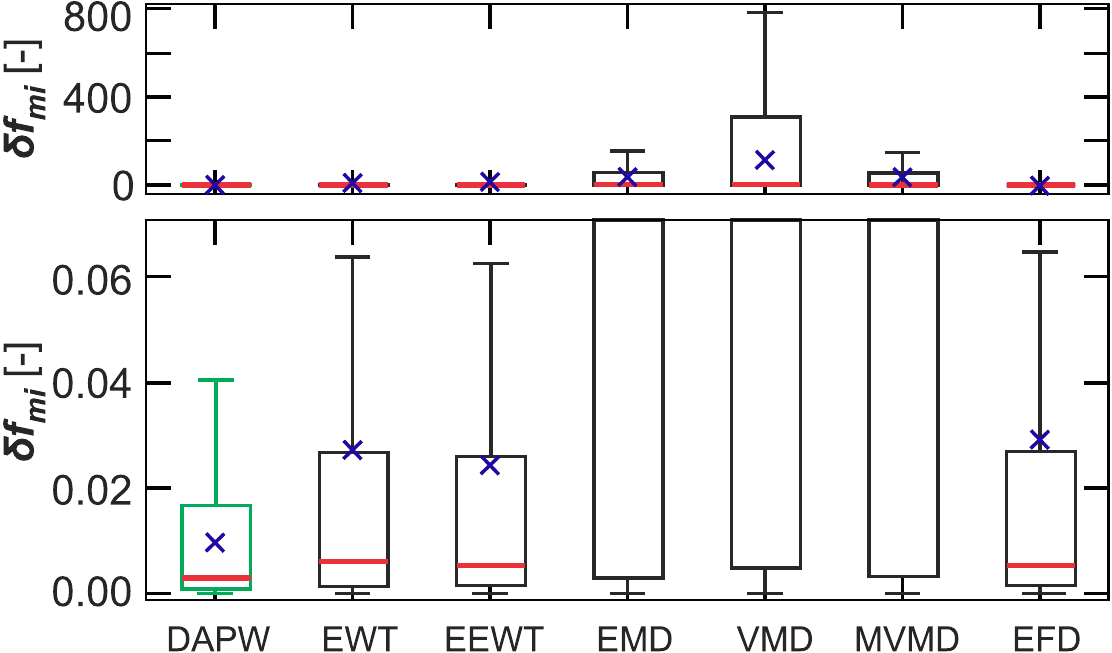}
		\caption{$\N$=3}
	\end{subfigure}
	\begin{subfigure}{0.5\textwidth}
		\centering
		\includegraphics[width=.6\columnwidth]{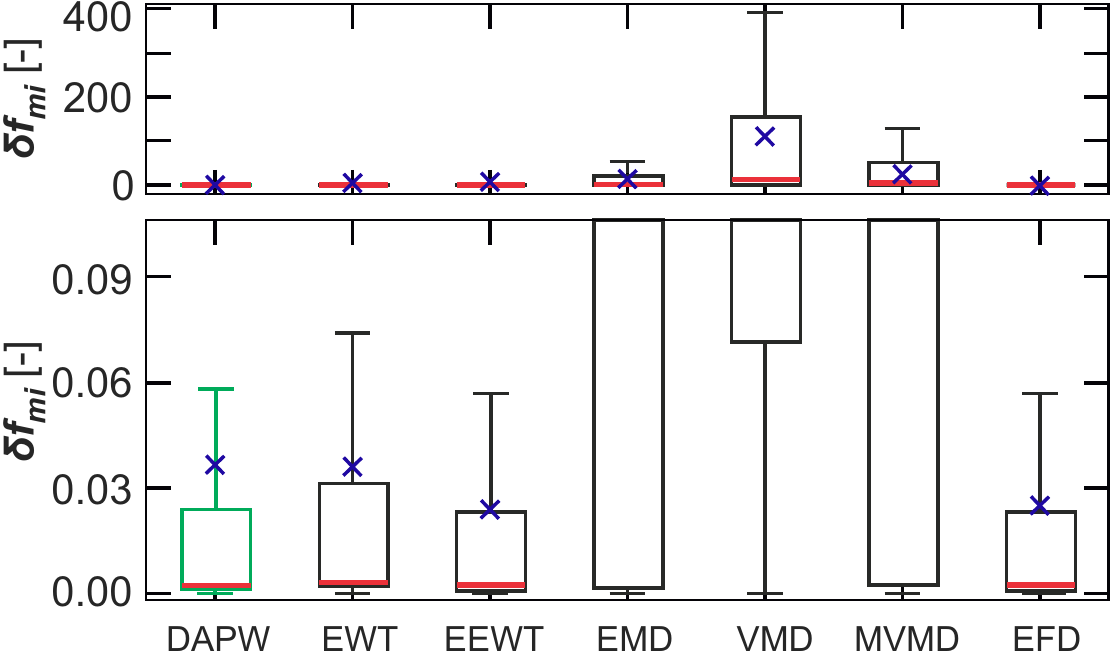}
		\caption{$\N$=4}
	\end{subfigure}
	\caption{The distribution of errors~$\dfmi$ for selected decomposition methods - numerical simulation studies}\label{Fig::result_ns_f}
\end{figure}

\begin{figure}[htb]\centering
	\centering
	\begin{subfigure}{0.5\textwidth}
		\centering
		\includegraphics[width=.6\columnwidth]{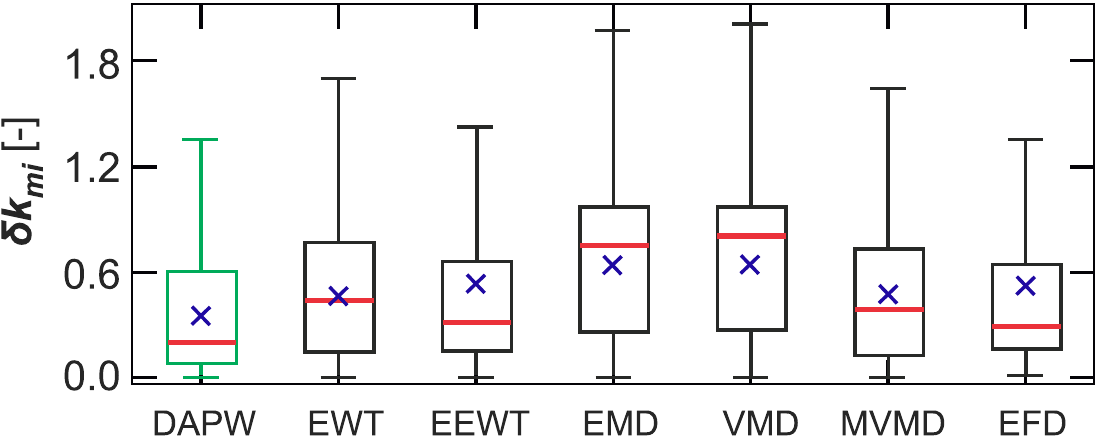}
		\caption{$\N$=2}
	\end{subfigure}
	\begin{subfigure}{0.5\textwidth}
		\centering
		\includegraphics[width=.6\columnwidth]{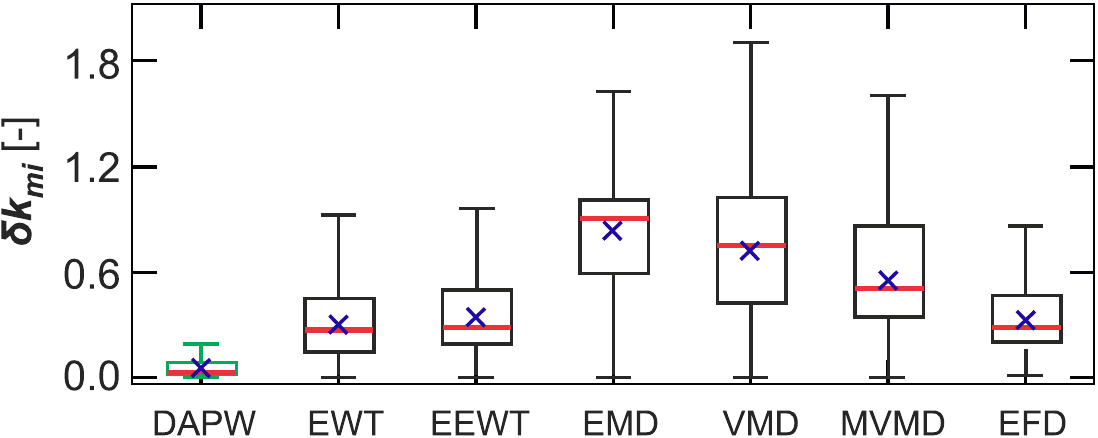}
		\caption{$\N$=3}
	\end{subfigure}
	\begin{subfigure}{0.5\textwidth}
		\centering
		\includegraphics[width=.6\columnwidth]{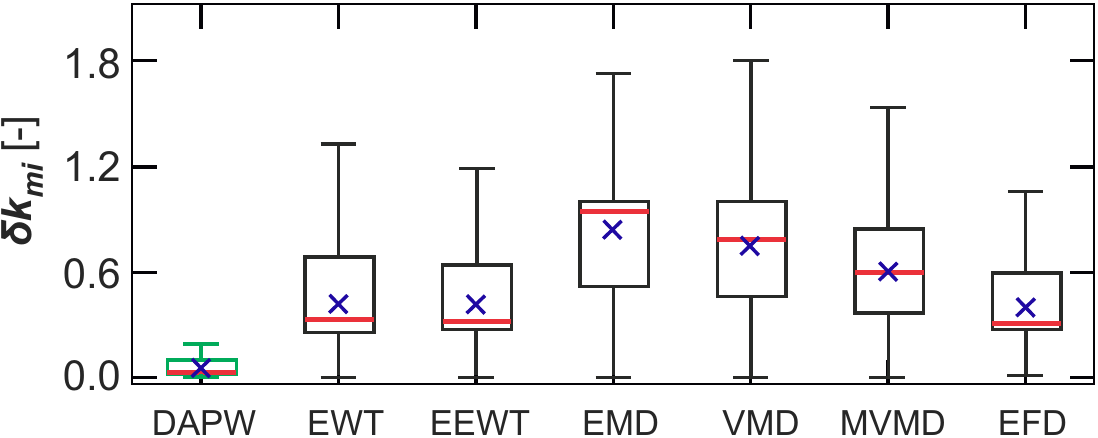}
		\caption{$\N$=4}
	\end{subfigure}
	\caption{The distribution of errors~$\dkmi$ for selected decomposition methods - numerical simulation studies}\label{Fig::result_ns_k}
\end{figure}

Fig.~\ref{Fig::result_ns_f} and Fig.~\ref{Fig::result_ns_k} show the statistical assessment of determined errors~$\dfmi$ and~$\dkmi$ in the form of ``box--plots" for results obtained from numerical simulation studies.

\begin{figure}[htb]\centering
	\centering
	\begin{subfigure}{0.5\textwidth}
		\centering
		\includegraphics[width=.6\columnwidth]{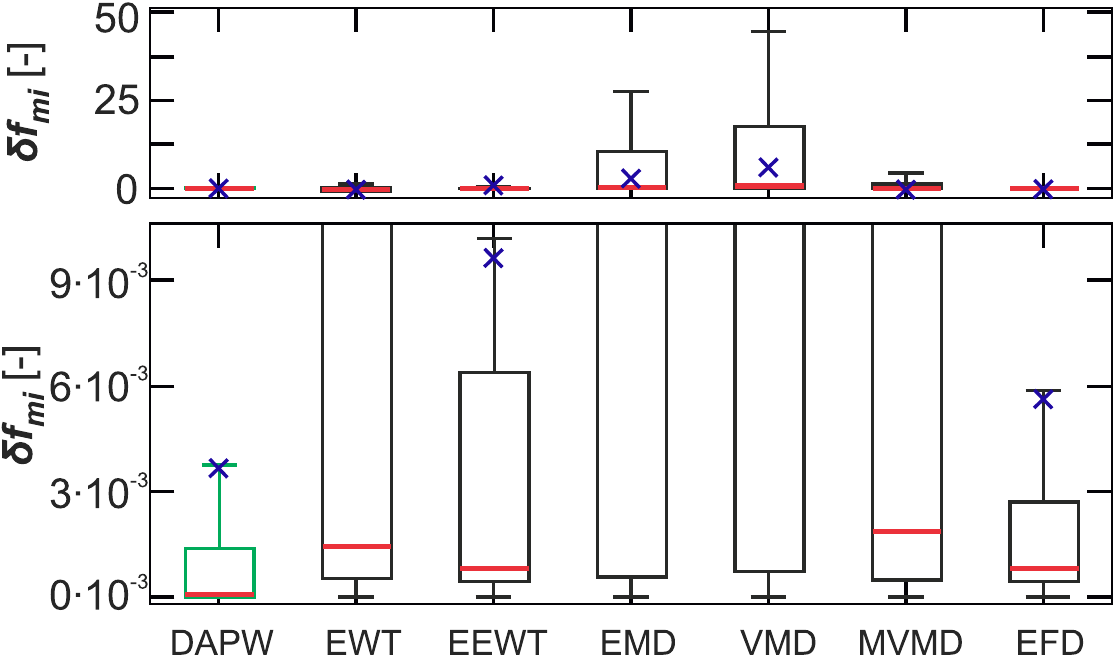}
		\caption{$\N$=2}
	\end{subfigure}
	\begin{subfigure}{0.5\textwidth}
		\centering
		\includegraphics[width=.6\columnwidth]{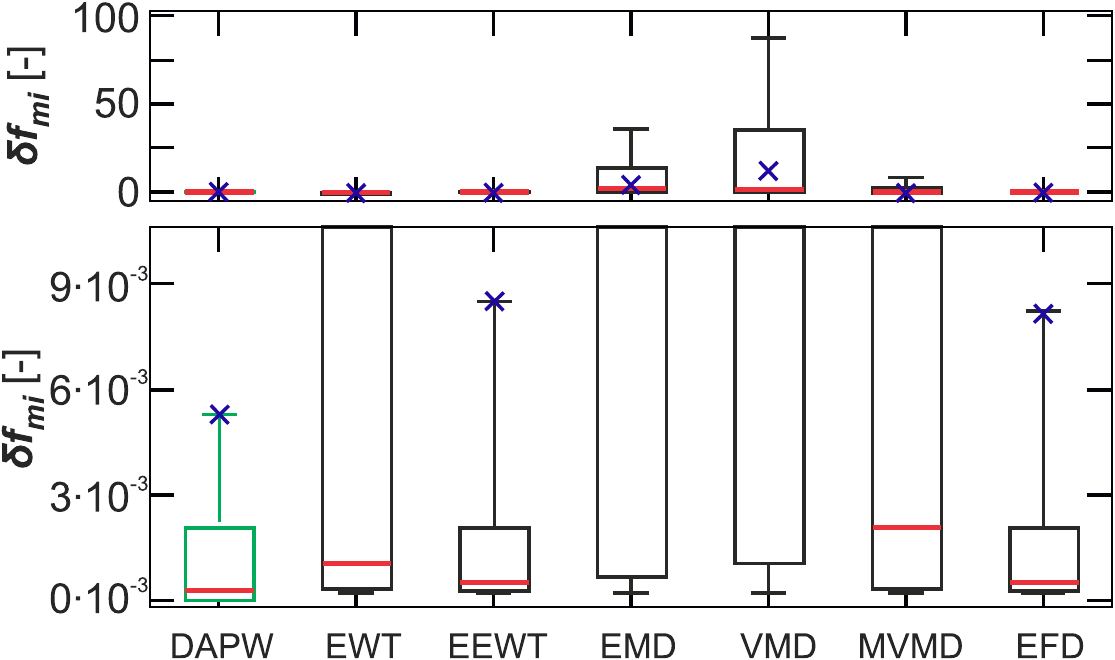}
		\caption{$\N$=3}
	\end{subfigure}
	\begin{subfigure}{0.5\textwidth}
		\centering
		\includegraphics[width=.6\columnwidth]{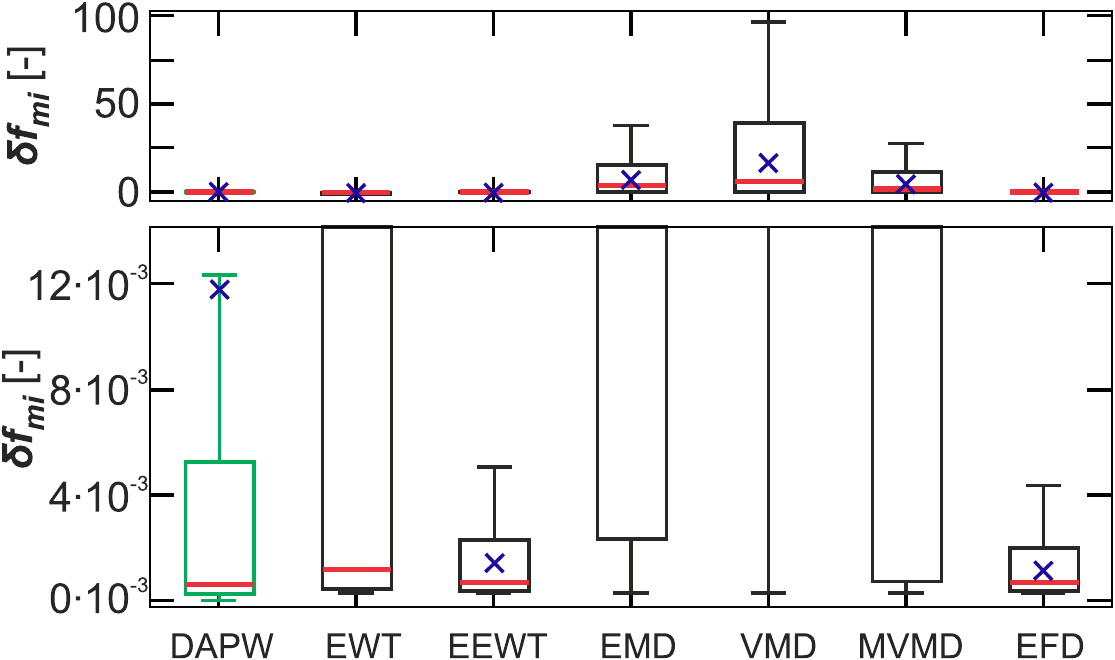}
		\caption{$\N$=4}
	\end{subfigure}
	\caption{The distribution of errors~$\dfmi$ for selected decomposition methods - laboratory experimental studies}\label{Fig::result_es_f}
\end{figure}

\begin{figure}[!]\centering
	\centering
	\begin{subfigure}{0.5\textwidth}
		\centering
		\includegraphics[width=.6\columnwidth]{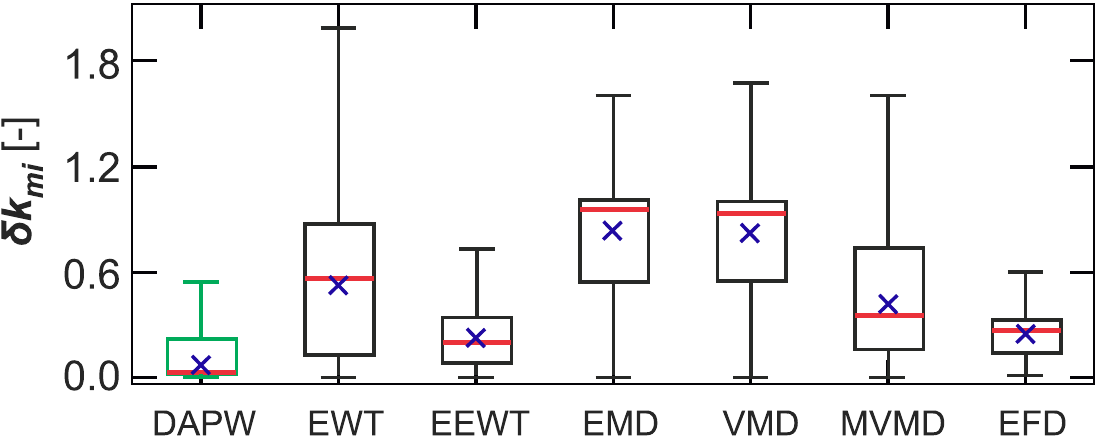}
		\caption{$\N$=2}
	\end{subfigure}
	\begin{subfigure}{0.5\textwidth}
		\centering
		\includegraphics[width=.6\columnwidth]{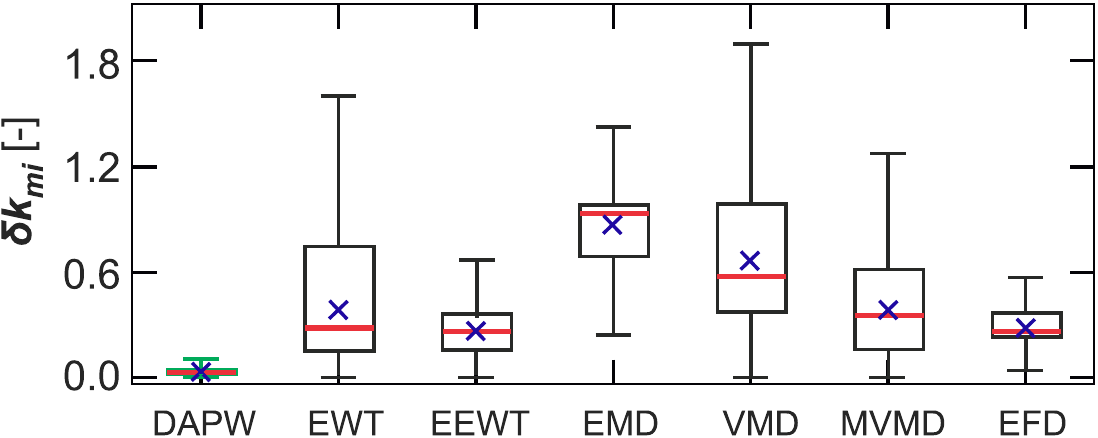}
		\caption{$\N$=3}
	\end{subfigure}
	\begin{subfigure}{0.5\textwidth}
		\centering
		\includegraphics[width=.6\columnwidth]{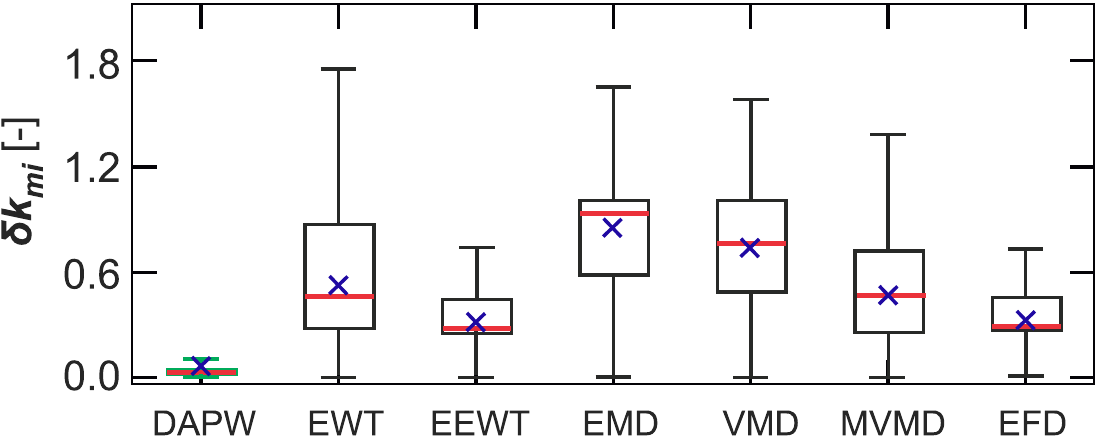}
		\caption{$\N$=4}
	\end{subfigure}
	\caption{The distribution of errors~$\dkmi$ for selected decomposition methods - laboratory experimental studies}\label{Fig::result_es_k}
\end{figure}

Fig.~\ref{Fig::result_es_f} and Fig.~\ref{Fig::result_es_k} show the statistical assessment of determined errors~$\dfmi$ and~$\dkmi$ in the form of ``box--plots" for results obtained from laboratory experimental studies.

Analyzing the distribution of errors presented in Figs.~\ref{Fig::result_ns_f}--\ref{Fig::result_es_k}, it can be seen that for the proposed approach, the smallest errors~$\dfmi$ and~$\dkmi$ are obtained. Figs.~\ref{Fig::result_ns_f} and~\ref{Fig::result_es_f} show that the average and median values of frequency estimation errors~$\dfmi$ of individual components are close to zero for the proposed approach. The smallest errors~$\dfmi$ obtained support the process of voltage fluctuations diagnostics focused on identification and initial recognition of disturbing loads. Figs.~\ref{Fig::result_ns_k} and~\ref{Fig::result_es_k} show that the mean and median values of errors in the estimation of amplitudes~$\dkmi$ of individual components are greater than for~$\dfmi$. However, errors~$\dkmi$ are acceptable, because for the correct localization (indication of the power supply point) of individual disturbing loads, it is necessary to estimate the amplitudes of individual component signals at individual points of the power grid while maintaining a constant value of relative error. Errors for decomposition methods other than the proposed method are related to the fact that they are based on decomposition into AM--FM modulated sinusoidal signals. However, voltage fluctuations of this type rarely occur in practice, so the diagnostic utility of methods other than the proposed one is lower.

It is worth noting that the proposed approach is based on functions that can be calculated according to the Cooley--Tukey algorithm when the number of samples in the measurement window is a power of 2. As a consequence, the complexity of the proposed solution is $\textbf{O}(n)$~\cite{hilb_comp}. Other decomposition methods that were used for comparison in this paper have a level of complexity of at least of $\textbf{O}(n^2)$, which significantly extends the calculation time when considering longer measurement windows used to consider slow voltage changes (e.g., time duration of measurement window equal to 10\unit{min}). For short time duration of windows, the time consumption of selected decomposition methods is comparable. However, it is worth noting that in order to correct recreation of the modulating signal with a frequency of up to 3$\fc$, it is necessary to use a sampling rate greater than 12\unit{kSa/s}, which in turn results in a large size of the input signals in the decomposition process for long measurement windows. 

\section{Conclusion}

Diagnostics of voltage fluctuations focused on selective identification and localization of sources of voltage fluctuations in the power grid, which change their operating state with a frequency of up to 3$\fc$, where $\fc$ is the power frequency, requires decomposition method with the smallest estimation errors~$\dfmi$ and~$\dkmi$ in the signal chain: ``demodulation -- decomposition -- statistical assessment of propagation". Therefore, the paper presents a new method of decomposition by approximation with pulse basis functions. Numerical simulation studies and laboratory experimental studies were carried out for the verification of accuracy of the proposed approach in its target application for the purposes of diagnostics of voltage fluctuations. The proposed approach is compared with other empirical decomposition methods available in the literature, intended for use for non--stationary and noisy signals, i.e., signals that are associated with real sources of voltage fluctuations, which can have a random nature. The analysis focused on step changes in voltage, considering the asymmetric operation of disturbing loads. Considering this type of modulating signal components is related to the fact that probably most sources of voltage fluctuations cause step voltage changes. The presented research results indicate that only the proposed approach allows for the correct identification of the frequency of individual disturbing loads in the power grid, which change their operating state with a frequency of up to 3$\fc$. In addition, the proposed approach is characterized by low computational complexity in the case of selection of the appropriate number of samples compared to other considered methods of empirical decomposition, which supports the process of implementation of the proposed approach in measuring and recording equipment with limited performance.

% References

\bibliographystyle{Bibliography/IEEEtranTIE}
\bibliography{Bibliography/IEEEabrv,Bibliography/BIB_1x-TIE-2xxx} %IEEEabrv instead of IEEEfull

\begin{thebibliography}{10}
\providecommand{\url}[1]{#1}
\csname url@samestyle\endcsname
\providecommand{\newblock}{\relax}
\providecommand{\bibinfo}[2]{#2}
\providecommand{\BIBentrySTDinterwordspacing}{\spaceskip=0pt\relax}
\providecommand{\BIBentryALTinterwordstretchfactor}{4}
\providecommand{\BIBentryALTinterwordspacing}{\spaceskip=\fontdimen2\font plus
\BIBentryALTinterwordstretchfactor\fontdimen3\font minus
  \fontdimen4\font\relax}
\providecommand{\BIBforeignlanguage}[2]{{%
\expandafter\ifx\csname l@#1\endcsname\relax
\typeout{** WARNING: IEEEtran.bst: No hyphenation pattern has been}%
\typeout{** loaded for the language `#1'. Using the pattern for}%
\typeout{** the default language instead.}%
\else
\language=\csname l@#1\endcsname
\fi
#2}}
\providecommand{\BIBdecl}{\relax}
\BIBdecl

\bibitem{pq_st_50160}
\emph{Voltage characteristics of electricity supplied by public electricity
  networks}, Standard {EN} 50160:2010/A2:2019, 2019.

\bibitem{tii_pqd_1}
O.~P. Mahela, B.~Khan, H.~H. Alhelou, and P.~Siano, ``Power quality assessment
  and event detection in distribution network with wind energy penetration
  using stockwell transform and fuzzy clustering,'' \emph{{IEEE} Trans. Ind.
  Informat.}, vol.~16, no.~11, pp. 6922--6932, 2020.

\bibitem{decomp}
R.~S. {H.}, S.~R. {Mohanty}, N.~{Kishor}, and A.~T. {K.}, ``Real-time
  implementation of signal processing techniques for disturbances detection,''
  \emph{{IEEE} Trans. Ind. Electron.}, vol.~66, no.~5, pp. 3550--3560, May.
  2019.

\bibitem{tie_pqd_1}
R.~Narayanaswami, D.~Sundaresan, and V.~Ranjan~Prema, ``The mystery curve: A
  signal processing based power quality disturbance detection,'' \emph{{IEEE}
  Trans. Ind. Electron.}, vol.~68, no.~10, pp. 10\,078--10\,086, 2021.

\bibitem{tie_pqd_2}
Q.~Tang, W.~Qiu, and Y.~Zhou, ``Classification of complex power quality
  disturbances using optimized s-transform and kernel svm,'' \emph{{IEEE}
  Trans. Ind. Electron.}, vol.~67, no.~11, pp. 9715--9723, 2020.

\bibitem{RaportWN}
``{6th CEER Benchmarking Report on all the Quality of Electricity and Gas
  Supply 2016},'' https://www.ceer.eu/, 2016.

\bibitem{flicker_ichqp}
M.~Michalski and G.~Wiczynski, ``Flicker dependency on voltage fluctuation at
  frequencies greater than power frequency,'' in \emph{20th ICHQP}, pp. 1--5,
  2022.

\bibitem{gnacinski_tec_2019}
P.~Gnacinski, M.~Peplinski, L.~Murawski, and A.~Szelezinski, ``Vibration of
  induction machine supplied with voltage containing subharmonics and
  interharmonics,'' \emph{{IEEE} Trans. Energy Convers.}, vol.~34, no.~4, pp.
  1928--1937, 2019.

\bibitem{granica150_1}
G.~{Wiczynski}, ``Sectional approximation of the flickermeter transformation
  characteristic for a~sinusoidal modulating signal,'' \emph{{IEEE} Trans.
  Instrum. Meas.}, vol.~57, no.~10, pp. 2355--2363, Oct. 2008.

\bibitem{granica150_2}
G.~{Wiczynski}, ``Simple model of flickermeter signal chain for deformed
  modulating signals,'' \emph{{IEEE} Trans. Power Del.}, vol.~23, no.~4, pp.
  1743--1748, Oct. 2008.

\bibitem{moje_TPD}
P.~Kuwalek, ``Estimation of parameters associated with individual sources of
  voltage fluctuations,'' \emph{{IEEE} Trans. Power Del.}, vol.~36, no.~1, pp.
  351--361, 2021.

\bibitem{elektr1}
H.~Karawia, M.~Mahmoud, and M.~Sami, ``Flicker in distribution networks due to
  photovoltaic systems,'' \emph{CIRED - Open Access Proc. J.}, vol. 2017, pp.
  647--649, Oct. 2017.

\bibitem{def_wsk}
G.~{Wiczynski}, ``Analysis of voltage fluctuations in power networks,''
  \emph{{IEEE} Trans. Instrum. Meas.}, vol.~57, no.~11, pp. 2655--2664, Nov.
  2008.

\bibitem{moje_En}
P.~Kuwalek, ``Selective identification and localization of voltage fluctuation
  sources in power grids,'' \emph{Energies}, vol.~14, no.~20, 2021.

\bibitem{single_point_1}
Z.~{Hanzelka}, ``Single point methods for location of electromagnetic
  disturbances in power system,'' \emph{Przeglad Elektrotechniczny}, vol.~91,
  no.~6, pp. 1--5, 2015.

\bibitem{single_point_5}
G.~A. {Senderovich} and A.~V. {Diachenko}, ``A method for determining location
  of voltage fluctuations source in electric grid,'' \emph{Electrical
  engineering and electromechanics}, no.~3, pp. 58--61, 2016.

\bibitem{multi_point_1}
N.~{Eghtedarpour}, E.~{Farjah}, and A.~{Khayatian}, ``Intelligent
  identification of flicker source in distribution systems,'' \emph{IET Gener.
  Transm. Dis.}, vol.~4, no.~9, pp. 1016--1027, 2010.

\bibitem{lok_wicz}
G.~{Wiczynski}, ``Voltage-fluctuation-based identification of noxious loads in
  power network,'' \emph{{IEEE} Trans. Instrum. Meas.}, vol.~58, no.~8, pp.
  2893--2898, 2009.

\bibitem{decomp_limit}
P.~Kuwalek, ``Decomposition problem in process of selective identification and
  localization of voltage fluctuation sources in power grids,'' in \emph{20th
  ICHQP}, pp. 1--6, 2022.

\bibitem{multi_point_3}
J.~J. Inamdar and K.~I. Annapoorani, ``A review of methods employed to identify
  flicker producing sources,'' \emph{TELKOMNIKA Telecomm. Comp. Electron. and
  Control}, vol.~16, pp. 465--480, 2018.

\bibitem{multi_point_2}
A.~Dejamkhooy, A.~Dastfan, and A.~Ahmadyfard, ``Source detection and
  propagation of equal frequency voltage flicker in nonradial power system,''
  \emph{Turkish Journal of Electrical Engineering and Computer Sciences},
  vol.~24, pp. 1351--1370, 2016.

\bibitem{sig_process}
J.~Semmlow, \emph{Circuits, Signals, and Systems for Bioengineers (Third
  Edition)}.\hskip 1em plus 0.5em minus 0.4em\relax Academic Press, 2018.

\bibitem{bien_am_fm}
K.~Duda, A.~Bien, M.~Szyper, and T.~Zielinski, ``Analysis of voltage
  disturbances caused by simultaneous amplitude and phase modulation in
  electric power network,'' in \emph{11th ICHQP}, pp. 199 -- 204, Oct. 2004.

\bibitem{moje_TIE}
P.~Kuwalek, ``{\color{blue}\MakeUppercase{AM}\color{black}} modulation signal
  estimation allowing further research on sources of voltage fluctuations,''
  \emph{{IEEE} Trans. Ind. Electron.}, vol.~67, no.~8, pp. 6937--6945, 2020.

\bibitem{ewt}
J.~{Gilles}, ``Empirical wavelet transform,'' \emph{{IEEE} Trans. Signal
  Process.}, vol.~61, no.~16, pp. 3999--4010, Aug. 2013.

\bibitem{eewt}
Y.~Hu, F.~Li, H.-G. Li, and C.~Liu, ``An enhanced empirical wavelet transform
  for noisy and non-stationary signal processing,'' \emph{Dig. Signal
  Process.}, vol.~60, pp. 220--229, Jan. 2017.

\bibitem{emd}
G.~Rilling, P.~Flandrin, and P.~Goncalves, ``On empirical mode decomposition
  and its algorithms,'' in \emph{IEEE-EURASIP Workshop on Nonlinear Signal and
  Image Processing}, pp. 8--11, 2003.

\bibitem{vmd}
K.~Dragomiretskiy and D.~Zosso, ``Variational mode decomposition,''
  \emph{{IEEE} Trans. Signal Process.}, vol.~62, no.~3, pp. 531--544, 2014.

\bibitem{mvmd}
N.~u. Rehman and H.~Aftab, ``Multivariate variational mode decomposition,''
  \emph{{IEEE} Trans. Signal Process.}, vol.~67, no.~23, pp. 6039--6052, 2019.

\bibitem{efd}
P.~Singh, S.~Joshi, R.~Patney, and K.~Saha, ``The {\color{blue}
  \MakeUppercase{f}ourier \color{black}} decomposition method for nonlinear and
  nonstationary time series analysis,'' \emph{Proc. R. Soc. A.}, vol. 473,
  2017.

\bibitem{f_measure}
Q.~{Lin} and Y.~{Shao}, ``A novel normalization method for autocorrelation
  funkction for pitch detection and for speech activity detection,'' in
  \emph{Int. Conf. on Interspeech}, pp. 2097--2101, 2018.

\bibitem{hilb_comp}
V.~Madisetti, \emph{The Digital Signal Processing Handbook}, 2nd~ed.\hskip 1em
  plus 0.5em minus 0.4em\relax Boca Raton, FL, USA: CRC Press, Inc., 2009.

\end{thebibliography}

%\vspace{-1cm}
%\begin{IEEEbiography}[{\includegraphics[width=1in,height=1.25in,clip,keepaspectratio]{Figures/AUTHOR_KUWALEK.jpg}}]
%	{P. Kuwa{\l{}}ek} received the B.Sc. degree in mathematics, M.Sc. Eng. in electrical engineering, and Ph.D. degrees in automation, electronic and electrical engineering from the Poznan University of Technology, Poznan, Poland, in 2018, 2018 and 2022, respectively.
%	
%	He is currently with the Division of Metrology, Electronics and Lighting Engineering at Poznan University of Technology. He has been the leader or the coordinator/ specialist in signal processing in many projects, including research and development projects. His current research interests include power quality evaluation and signal processing. 
%\end{IEEEbiography}

\end{document}